\documentclass[aps, prapplied, reprint, superscriptaddress]{revtex4-1}
\usepackage{amssymb}
\usepackage{amsmath}
\usepackage{graphicx}
\usepackage{dcolumn}
\usepackage{bm}

\begin{document}

\title{ Ballistic hot-electron transport in a quantum Hall edge channel
defined by a double gate}
\author{Shunya Akiyama}
\author{Taichi Hirasawa}
\author{Yuya Sato}
\affiliation{Department of Physics, Tokyo Institute of Technology, 2-12-1 Ookayama,
Meguro, Tokyo, 152-8551, Japan.}
\author{Takafumi Akiho}
\author{Koji Muraki}
\affiliation{NTT Basic Research Laboratories, NTT Corporation, 3-1 Morinosato-Wakamiya,
Atsugi 243-0198, Japan.}
\author{Toshimasa Fujisawa}
\email{fujisawa@phys.titech.ac.jp}
\affiliation{Department of Physics, Tokyo Institute of Technology, 2-12-1 Ookayama,
Meguro, Tokyo, 152-8551, Japan.}

\begin{abstract}
Ballistic transport of hot electrons in a quantum Hall edge channel is
attractive for studying electronic analog of quantum optics, where the edge
potential profile is an important parameter that governs the charge velocity
and scattering by longitudinal-optical (LO) phonons. Here we use a parallel
double gate to control the electric field of the edge potential, and
investigate the ballistic length of the channel by using hot-electron
spectroscopy. The ballistic length is significantly enhanced by reducing the
LO phonon scattering rate in the tailored potential.
\end{abstract}

\date{\today }
\maketitle

Scattering of conduction electrons is crucial for studying novel quantum
transport \cite{BookNazarov}. Scattering by impurities can be suppressed in
high-mobility heterostructures, where cold electrons show ballistic electron
transport with mean free path beyond 100 $\mu $m \cite%
{BookBeenakker,HirayamaBallistic}. The impurity scattering can also be
evaded in quantum Hall edge channels under a strong magnetic field, where
cold electrons travel along an equipotential line at the chemical potential
even in the presence of impurities \cite{BookEzawa,HalperinPRB1982}.
Successful realizations of fermionic interferometers \cite%
{MZI-Nature2013,HOM-NatComm2015} and quantum tomography \cite%
{QTomography2019} are promising for quantum information channels. While
single-particle excitations relax into many-body excitations in a short
distance as a result of electron-electron scattering with nearby channels 
\cite{TaubertPRB2011,ItohPRL2018,AugerQH2019}, long coherence length of
sub-millimeters was recently reported by tailoring the copropagating channel 
\cite{RelaxQH21-2019,QH-MZI21-2019}. On the other hand, the
electron-electron scattering can also be suppressed significantly by
exciting the electron of interest far above the chemical potential, where
the hot electron in a soft edge potential of an AlGaAs/GaAs heterostructure
cannot satisfy the momentum conservation for the scattering \cite{OtaPRB2019}%
. Long ballistic transport over millimeters is quite attractive particularly
for studying electronic analog of quantum optics \cite%
{JohnsonPRL2018,ReviewElecQO}. For example, pure single-electron excitation
with well-defined energy can be generated from a dynamic quantum dot \cite%
{FletcherPRL2013}. The hot electron travels at velocity determined by the
edge potential \cite{KataokaPRL2016}. The energy and time uncertainties of a
hot-electron wave packet can reach a level comparable to the quantum limit 
\cite{FletcherArXiv2019}. Channels for hot electrons can be tailored to
implement a beam-splitter and a charge detector \cite%
{FreiseFCS-arXiv2019,UbbelohdePartitioning-NatNano2014}. In these
experiments, the dominant scattering mechanism can be scattering by
longitudinal optical (LO) phonons. Particularly for long ballistic
transport, the phonon scattering should be suppressed by making the edge
potential less steep, i.e., by lowering the edge electric field \cite%
{JohnsonPRL2018,EmaryPRB2019,OtaPRB2019}. As the electric field also
influences the hot-electron velocity, it would be better to have tuning
capability of the electric field. In most of the previous studies, an
etching step covered with a flat gate was used to define edge channels \cite%
{KataokaPRL2016,JohnsonPRL2018}, so that little attention has been focused
on the active tuning of the electric field.

Here we investigate ballistic hot-electron transport and LO phonon
scattering in a tunable edge potential defined by two parallel fine gates
along the channel. The scattering length is investigated by employing
hot-electron spectroscopy with injector and detector point contacts. The
scattering rate shows a non-monotonic energy dependence in agreement with
electrostatic simulation. The double gate geometry can be used to tailor the
edge potential for tuning the scattering as well as the velocity.

As shown in Fig. 1(a), a double gate with G$_{1}$ and G$_{2}$ was patterned
on an AlGaAs/GaAs heterostructure. Large negative voltage $V_{\mathrm{G}1}$
on G$_{1}$ primary defines the right-moving edge channel under perpendicular
magnetic field $B$, and small negative voltage $V_{\mathrm{G}2}$ on G$_{2}$
modifies the edge potential. To see its effect, we performed electrostatic
calculations based on the Poisson equation by employing a two-dimensional
(2D) finite element method with realistic parameters for our device \cite%
{OtaPRB2019,Kamata2014-TLL,Hashisaka2017-TLL}. For simplicity, we used the
frozen surface model with a fixed charge density on the free surface \cite%
{DaviesLarkin-PRB1994}. The bulk region of the 2D electrons is regarded as a
grounded metal, as the energy of hot electrons in this work (30 - 100 meV)
is much higher than the Fermi energy of the 2D electrons ($\lesssim $ 10
meV). The boundary position of the metallic region representing the 2D
electron system was determined self-consistently in such a way that the
charge density on the metallic region becomes zero at the boundary position.

With large negative $V_{\mathrm{G}1}$ applied while keeping $V_{\mathrm{G}2}$
= 0, the potential energy $e\phi \left( x\right) $ along the transverse axis 
$x$ to the channel [see Fig. 1(a)] has a maximum underneath G$_{1}$ as shown
by the black dashed line in Fig. 1(b). In a high magnetic field, this
spatial profile can be regarded as that of the lowest Landau level (LLL).
Hot electrons travel on the left side of the maximum at single-particle
velocity $v_{\mathrm{hot}}=E/B$, where $E=-d\phi /dx$ is the electric field.
We focus on the LO phonon scattering within the LLL, where an electron emits
the LO phonon energy of $\varepsilon _{\mathrm{LO}}$ = 36 meV with a spatial
shift $d$ between the initial and final states \cite{EmaryPRB2019}. By
neglecting the non-linearity in the potential for simplicity, the LO phonon
scattering rate can be written as $\Gamma _{\mathrm{LO}}=\Gamma _{0}\exp
\left( -d^{2}/2\ell _{b}^{2}\right) $, which depends strongly on $d\simeq
\varepsilon _{\mathrm{LO}}/eE$. Here, $\ell _{b}=\sqrt{\hbar /eB}$ is the
magnetic length and $\Gamma _{0}$ is the form factor for the LO-phonon
scattering. The scattering length $\ell _{\mathrm{LO}}=$ $v_{\mathrm{hot}%
}/\Gamma _{\mathrm{LO}}$ can be enhanced significantly by decreasing $\Gamma
_{\mathrm{LO}}$ under the condition of $d\gtrsim \ell _{b}$, and thus by
decreasing $E$. This can be done by applying negative $V_{\mathrm{G}2}$; as
shown in Fig. 1(b), the slope of the potential energy $e\phi $, $E$, is
reduced in a particular energy range that depends on $V_{\mathrm{G}2}$.
Figure 1(c) shows the potential energy $e\phi $ dependence of the calculated
electric field $E$, where a clear dip in $E$ develops as negative $V_{%
\mathrm{G}2}$ is applied. The impact on $\ell _{\mathrm{LO}}$ at a
representative magnetic field of $B$ = 9 T is shown by the subsidiary scale
(arbitrary unit). While this crude simulation can be used as a qualitative
guide, huge enhancement of $\ell _{\mathrm{LO}}$ is suggested with a small
reduction of $E$. We investigate such potential tuning by measuring $\ell _{%
\mathrm{LO}}$ in this work. We mainly focus on the LO phonon scattering
within the\ LLL, while other mechanisms like electron-electron scattering
and inter-Landau-level tunneling are important in some other conditions \cite%
{JohnsonPRL2018,OtaPRB2019,Komiyama1992-ILL}.

The LO phonon scattering length $\ell _{\mathrm{LO}}$ is experimentally
evaluated by using hot electron spectroscopy. As shown in the device
geometry of Fig. 1(a) and the energy diagram of Fig. 2(a), hot electrons are
injected from emitter E, which is biased at $eV_{\mathrm{E}}$ = 10 - 200
meV, through the injector point contact defined with gate G$_{\mathrm{inj}}$%
. After traveling along the edge of the base (B) region, the electrons are
analyzed with the detector point contact defined with gate G$_{\mathrm{det}}$%
. Hot electrons with energy greater than the barrier height $\varepsilon
_{\det }$ are introduced to the collector, while other electrons are
reflected and lead to the base ohmic contact. The energy distribution
function is obtained by measuring the collector current $I_{\mathrm{col}}$
or the reflected current $I_{\mathrm{ref}}$ as a function of gate voltage $%
V_{\mathrm{det}}$ on G$_{\mathrm{det}}$ which determines $\varepsilon _{\det
}$.

Since both G$_{1}$ and G$_{2}$ constitute the point contact in conjunction
with G$_{\mathrm{det}}$ in our device as shown in Fig. 1(a), the potential
modulation with G$_{2}$ also influences the detector characteristics. Figure
1(d) shows the approximate potential profile $\phi \left( x^{\prime }\right) 
$ along the transverse axis $x^{\prime }$ around the detector point contact
[see Fig. 1(a)], where the wedge-shaped G$_{\mathrm{det}}$ is crudely
approximated by an infinitely wide gate so as to fit with the 2D simulation.
The potential minimum marked by the open circles corresponds to the barrier
height or the saddle-point energy $\varepsilon _{\det }$ in the actual
device. The simulation shows that $\varepsilon _{\det }$ increases with
decreasing $V_{\mathrm{det}}$. Figure 1(e) shows the relation between $%
\varepsilon _{\det }$ and $V_{\mathrm{det}}$, where one can see that the
slope, or the lever-arm factor $\alpha =\Delta \varepsilon _{\det }/\Delta
V_{\mathrm{det}}$, changes abruptly at a particular point that depends on $%
V_{\mathrm{G}2}$. The $\varepsilon _{\det }$ value at this bending point in
Fig. 1(e) is close to the $e\phi $ value at the dip in Fig. 1(c). We also
use this detector characteristics to confirm the double gate action.

\begin{figure}[tbp]
\begin{center}
\includegraphics[width = 3.15in]{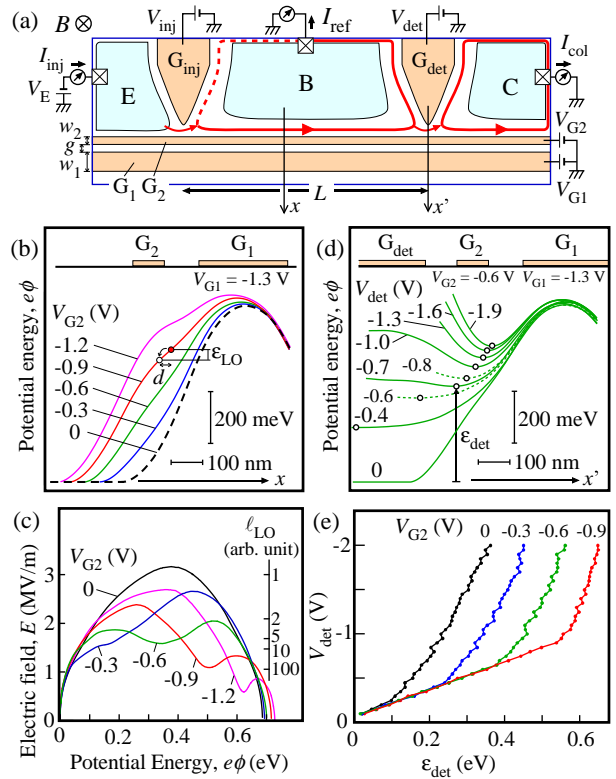}
\end{center}
\caption{ (a) An experimental setup for hot-electron spectroscopy on an edge
channel formed by a double gate, G$_{1}$ and G$_{2}$. (b) Potential energy
profile $e\protect\phi \left( x\right) $ of the edge channel. The locations
of G$_{1}$ and G$_{2}$ in the $x$ axis are shown on the top. (c) The
electric field $E=-d\protect\phi /dx$ as a function of the potential energy $%
e\protect\phi $. The LO-phonon scattering length $\ell _{\mathrm{LO}}$ in an
arbitrary unit for $B$ = 9 T is shown. (d) Approximate potential energy
profile $e\protect\phi \left( x^{\prime }\right) $ at the detector point
contact. The locations of G$_{1}$, G$_{2}$, and G$_{\det }$ in the $%
x^{\prime }$ axis are shown. (e) The detector energy $\protect\varepsilon %
_{\det }$ (in the horizontal axis) as a function of $V_{\mathrm{det}}$ (in
the vertical axis) for several $V_{\mathrm{G}2}$ values. Plots in (b)-(e)
were obtained by numerically solving the Poisson equation with a 2D
finite-element method by assuming translational invariance along the
transport direction. The surface potential is fixed at $eV_{g}$ + 0.6 eV in
the gated region, but surface electron density is fixed at 6$\times $10$%
^{11} $ cm$^{-2}$ in the ungated region. A uniform donor density of 9$\times 
$10$^{11}$ cm$^{-2}$ is assumed. Experimental geometry with $w_{1}$ = 280
nm, $w_{2}$ = 100 nm, and $g$ = 100 nm, were used in the simulation. }
\end{figure}

The double gate devices with G$_{1}$ of width $w_{1}$ = 280 nm and G$_{2}$
of $w_{2}$ = 100 nm separated by gap $g$ = 100 nm were fabricated by
patterning a Ti layer on an AlGaAs/GaAs heterostructure. The two-dimensional
electron system located at 100 nm below the surface has an electron density
of 2.7$\times $10$^{11}$ cm$^{-2}$ and low-temperature mobility of about 10$%
^{6}$ cm$^{2}$/Vs. Under the perpendicular magnetic field of $B$ = 9 T,
appropriate gate voltages were applied to form injection and detection point
contacts as illustrated in Fig. 1(a). While the bulk is conductive with
Landau filling factor $\nu =$ 1.3, this should not influence the
hot-electron transport that is well isolated from the cold electrons in the
edge and the bulk \cite{OtaPRB2019}. We investigated two devices with the
distance $L$ = 2 and 4 $\mu $m between the injector and detector point
contacts, and both devices show similar results. The following data was
taken with the $L$ = 2 $\mu $m device at 1.5 K.

\begin{figure}[tbp]
\begin{center}
\includegraphics[width = 3in]{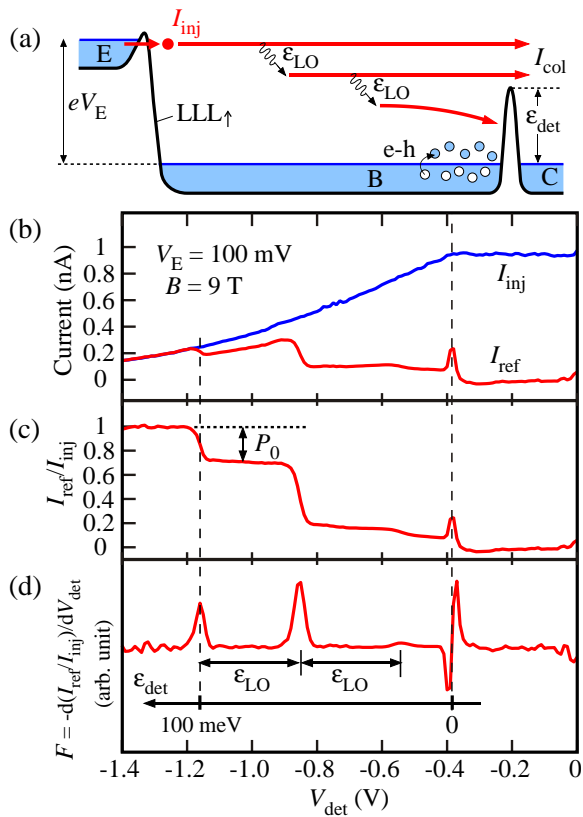}
\end{center}
\caption{(a) Schematic energy diagram of the hot-electron spectroscopy. (b) $%
I_{\mathrm{ref}}$ and $I_{\mathrm{inj}}$, (c) $I_{\mathrm{ref}}/I_{\mathrm{%
inj}}$, and (d) $F=-d(I_{\mathrm{ref}}/I_{\mathrm{inj}})/dV_{\mathrm{\det }}$
as a function of $V_{\mathrm{det}}$ for $L=$ 2 $\protect\mu $m device. The
equispaced peaks in (d) represent the ballistic transport (the leftmost
peak) and its phonon replicas. The step height $P_{0}$ in (c) measures the
probability of ballistic transport. }
\end{figure}

Figure 2(b) shows a representative data of injection current $I_{\mathrm{inj}%
}$ and reflected current $I_{\mathrm{ref}}$ as a function of the detector
gate voltage $V_{\mathrm{det}}$ at $eV_{\mathrm{E}}$ = 100 meV. We adjusted $%
V_{\mathrm{inj}}$ to provide $I_{\mathrm{inj}}\simeq $ 1 nA at $V_{\mathrm{%
det}}$ = 0, and swept $V_{\mathrm{det}}$ to measure the current profile.
While the application of negative $V_{\mathrm{det}}$ slightly reduces the
injection current due to the electrostatic coupling, the reflected current $%
I_{\mathrm{ref}}$ increases and eventually reaches the value identical to $%
I_{\mathrm{inj}}$ when a large barrier is formed at the detector. We
confirmed that the total current is always conserved, i.e., $I_{\mathrm{inj}%
}=I_{col}+I_{\mathrm{ref}}$ (not shown). The step-like features in
the normalized current $I_{\mathrm{ref}}/I_{\mathrm{inj}}$ in Fig. 2(c) are
associated with the emission of LO phonons with $\varepsilon _{\mathrm{LO}}$
= 36 meV. The height of the first step, $P_{0}$, in $I_{\mathrm{ref}}/I_{%
\mathrm{inj}}$ measures the probability (fraction) that the injected
electron reaches the collector ballistically without emitting LO phonons.
This is evaluated with scattering length $\ell _{\mathrm{LO}}$ defined by
the relation $P_{0}=\exp \left( -L/\ell _{\mathrm{LO}}\right) $. The
derivative $F=-d(I_{\mathrm{ref}}/I_{\mathrm{inj}})/dV_{\mathrm{det}}$ in
Fig. 2(d) is proportional to the energy distribution function of the edge
state \cite{OtaPRB2019}. The three peaks at $V_{\mathrm{det}}\simeq $ $-$%
1.16, $-$0.84, and $-$0.54 V are equally spaced with the separation ($\sim $
0.31 V) corresponding to $\varepsilon _{\mathrm{LO}}$ = 36 meV in energy.
This indicates linear $\varepsilon _{\det }-V_{\mathrm{det}}$ dependence in
this energy range with a lever-arm factor $\alpha =\Delta \varepsilon _{\det
}/\Delta V_{\mathrm{det}}\simeq $ 0.12$e$. The peak in $I_{\mathrm{ref}}$\
(the dip and peak in $F$) at $V_{\mathrm{det}}\simeq -0.38$ V originates
from electrical noise in the collector, where the noise could be rectified
by the nonlinear characteristics of the point contact. Nevertheless, this
marks the condition where the detector barrier is effectively zero ($%
\varepsilon _{\det }=0$). Considering that the peak position at $V_{\mathrm{%
det}}\simeq $ $-$0.54 V corresponds to the energy of $eV_{\mathrm{E}%
}-2\varepsilon _{\mathrm{LO}}$ = 28 meV, slightly larger lever-arm factor $%
\alpha \simeq $ 0.18$e$ is suggested in the low energy range of $-0.54$ V $%
<V_{\mathrm{det}}<$ $-$0.38 V. The difference in the lever-arm factor
becomes larger when finite $V_{\mathrm{G}2}$ is applied as shown below.

Figure 3 shows color plots of $F$ as a function of $V_{\mathrm{det}}$ and
the injection energy $eV_{\mathrm{E}}$. The data taken at $V_{\mathrm{G}2}$
= 0 V [Fig. 3(a)] reveal several peaks, including the topmost ballistic
signal and its phonon replicas. The topmost peak follows the ballistic
condition ($\varepsilon _{\det }=eV_{\mathrm{E}}$), shown by the dashed line
by assuming linear dependence of $V_{\mathrm{det}}$ on $\varepsilon _{\det }$%
, for $V_{\mathrm{E}}>$ 60 mV. The amplitude of the ballistic peak decreases
with increasing $eV_{\mathrm{E}}$ as the LO phonon scattering rate increases
with the electric field $E$ at higher energy \cite{OtaPRB2019}. In contrast,
the topmost peak for $V_{\mathrm{E}}<$ 50 mV deviates from the ballistic
condition and disappears at $V_{\mathrm{E}}=V_{\mathrm{th}}\simeq $ 30 mV.
This is associated with electron-electron scattering which becomes more
significant at lower energy; indeed, the threshold energy $eV_{\mathrm{th}}$
is comparable to the value in our previous report \cite{OtaPRB2019}. In the
following, we focus on the LO phonon scattering observed at $V_{\mathrm{E}}>$
60 mV.

\begin{figure}[tbp]
\begin{center}
\includegraphics[width = 3.15in]{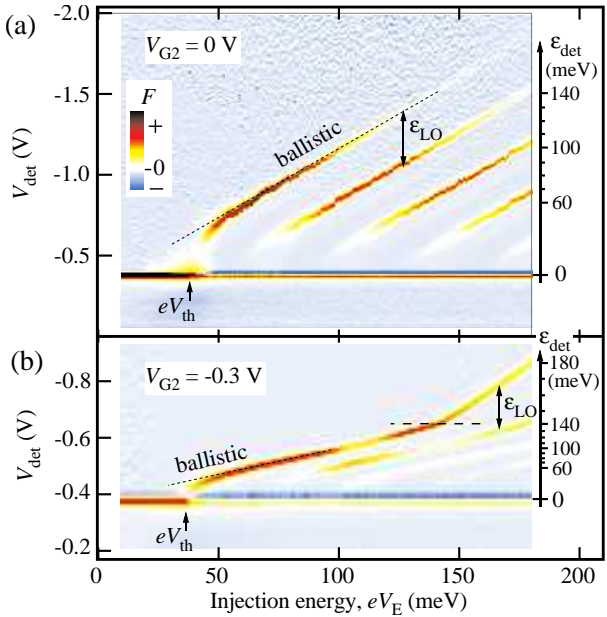}
\end{center}
\caption{ Representative hot-electron spectrum $F$ as a function of $V_{\det
}$ for various $eV_{E}$ taken at $V_{\mathrm{G2}}=$ 0 V in (a) and $V_{%
\mathrm{G2}}=$ $-$0.3 V in (b). The data was taken at $B$ = 9 T (the bulk
filling factor $\protect\nu =$ 1.3) and $V_{\mathrm{G1}}=$ $-$1.3 V. The
subsidiary scale of $\protect\varepsilon _{\det }$ is determined from the
ballistic peak position for $eV_{E}>$ 60 meV. }
\end{figure}

The hot-electron spectrum changes dramatically when $V_{\mathrm{G}2}=-0.3$ V
is applied as shown in Fig. 3(b). While the ballistic peak and its phonon
replicas are seen, their peak positions in $V_{\mathrm{det}}$ deviate
significantly from those in Fig. 3(a). Clear bending is seen at $V_{\mathrm{%
det}}\simeq $ -0.65 V marked by the horizontal dashed line. This is
qualitatively consistent with the electrostatic calculation of Fig. 1(e). By
assuming that the topmost ballistic peak follows $\varepsilon _{\det }=eV_{%
\mathrm{E}}$ for $V_{\mathrm{E}}>$ 60 mV, $\varepsilon _{\det }$ should
change with $V_{\mathrm{det}}$ in a non-linear way as shown by the
subsidiary scale in Fig. 3(b). Similar bending appears in a wide range of $%
V_{\mathrm{G}2}$ as summarized in Figs. 4(a) and 4(c), where the position of
the ballistic peak is plotted as a function of $eV_{\mathrm{E}}$ for several 
$V_{\mathrm{G}2}$ values in Fig. 4(a) and for several $V_{\mathrm{G}1}$
values in Fig. 4(c). The bending position changes with $V_{\mathrm{G}2}$ and 
$V_{\mathrm{G}1}$ in qualitative agreement with the electrostatic
calculation. Note that peak positions of the phonon-replicas also show
bending (not shown) at the same position in $V_{\mathrm{det}}$ as the
ballistic peak position, indicating that the bending originates in the
detector. This ensures that the edge potential is modulated with $V_{\mathrm{%
G}2}$.

\begin{figure}[tbp]
\begin{center}
\includegraphics[width = 3.15in]{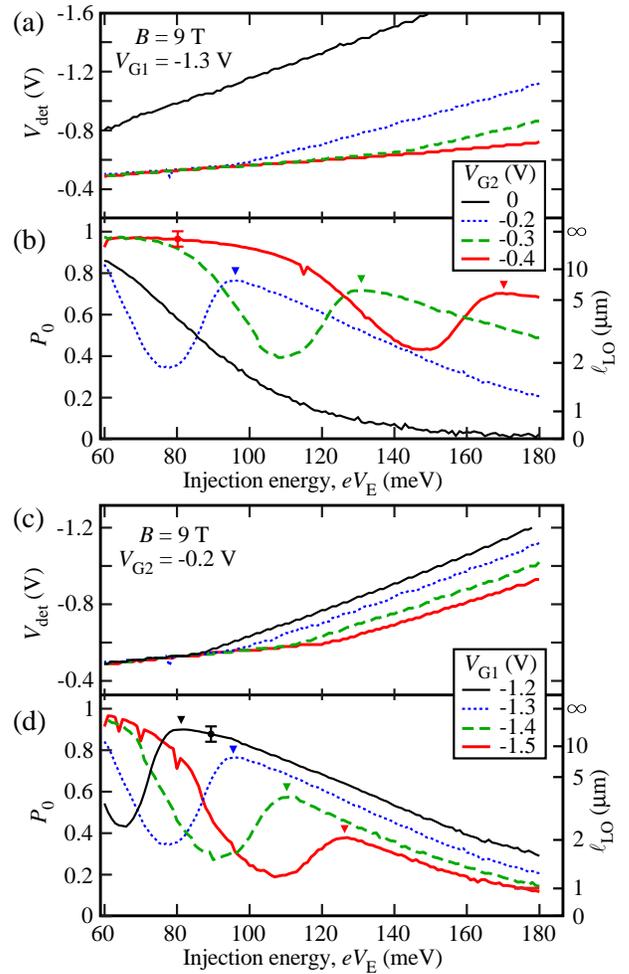}
\end{center}
\caption{ Injection energy ($eV_{\mathrm{E}}$) dependence of the ballistic
peak position in $V_{\det }$ in (a) and (c) and the ballistic transport
probability $P_{0}$ in (b) and (d) for several $V_{\mathrm{G2}}$ values at $%
V_{\mathrm{G1}}=$ -1.3 V in (a) and (b) and for several $V_{\mathrm{G1}}$
values at $V_{\mathrm{G2}}=$ -0.2 V in (c) and (d). The corresponding
scattering length $\ell _{\mathrm{LO}}$ is shown in the right scale of (b)
and (d). The triangles mark the enhancement of $P_{0}$ (and thus $\ell _{%
\mathrm{LO}}$) with the gentle edge potential. A representative error bar is
shown in (b) and (d).}
\end{figure}

The peak height (color) of the ballistic signal plotted in Fig. 3(b) shows a
complicated dependence on $V_{\mathrm{E}}$, with two maxima at $V_{\mathrm{E}%
}\simeq $ 70 and 130 mV. While the first maximum at $V_{\mathrm{E}}\simeq $
70 mV should be explained with the electron-electron scattering as in Fig.
3(a), the second maximum at $V_{\mathrm{E}}\simeq $ 130 mV close to the
bending position indicates the reduction of LO phonon scattering rate. The
step height $P_{0}$ in $I_{\mathrm{ref}}/I_{\mathrm{inj}}$ of Fig. 2(c),
which measures the probability of the ballistic transport over the channel
length $L$, is plotted as a function of the injection energy $eV_{\mathrm{E}}
$ ($>$ 60 meV) for several $V_{\mathrm{G}2}$ values in Fig. 4(b) and for
several $V_{\mathrm{G}1}$ values in Fig. 4(d). Corresponding scattering
length $\ell _{\mathrm{LO}}$ is shown on the right scale. While $P_{0}$ and
thus $\ell _{\mathrm{LO}}$ monotonically decrease with $eV_{\mathrm{E}}$ at $%
V_{\mathrm{G}2}=0$ in Fig. 4(b), non-monotonic dependence of $P_{0}$ with a
broad peak, marked by triangles, is seen at negative $V_{\mathrm{G}2}$. The
peak position moves to the high energy side by making $V_{\mathrm{G}2}$ more
negative. This behavior is qualitatively consistent with that of the dip
position in $E$ [Fig. 1(c)]. Similarly, the peak position moves to the high
energy side by making $V_{\mathrm{G}1}$ more negative in Fig. 4(b), which is
also consistent with the simulation (not shown). In this way, we have
successfully modified the edge potential and the LO phonon scattering length
with the double gate geometry.

The modulation of the edge potential may also influence the
electron-electron scattering in the low energy region ($eV_{\mathrm{E}}<$ 60
meV). Several competing effects are expected \cite{OtaPRB2019}. When the
potential slope is made less steep, the hot electron velocity decreases
toward the single-particle velocity at the Fermi energy, which would
contribute to increase the electron-electron scattering rate as the energy
and momentum conservation is met at equal velocities. On the other hand, the
hot electrons become spatially more separated from the Fermi sea, which
would work to decrease the scattering rate. While these two effects cancel
each other for a simplified potential profile $\phi \propto x^{\xi }$ with
an exponent $\xi $ and the unscreened Coulomb potential, the scattering rate
for the potential modified by $V_{\mathrm{G}2}$ in our device may change. In
addition, screening by the gate metal can suppress the electron-electron
scattering. Experimentally we did not see clear $V_{\mathrm{G}2}$ dependence
in the electron-electron scattering and the threshold energy $eV_{\mathrm{th}%
}$ at which the hot electron peak starts to appear in Fig. 3. Further
studies are encouraged on this effect.

In summary, we have demonstrated that the edge potential can be modulated
with a double gate geometry by measuring the LO phonon scattering length.
This scheme may be useful in extending the ballistic length and tuning the
charge velocity, and can be extended for fully depleted regions with more
than two gates. This is attractive for studying electronic analog of quantum
optics with hot electrons.

We thank Tomoaki Ota, Masayuki Hashisaka and Tokuro Hata for fruitful
discussions and supports. This work was supported by JSPS KAKENHI
(JP26247051, JP15H05854, JP17K18751, JP19H05603), and Nanotechnology
Platform Program of MEXT.

\end{document}